%&pdftex
\vsize=23.5truecm \hsize=17truecm
\baselineskip=0.5truecm \parindent=0truecm
\parskip=0.2cm \hfuzz=1truecm

\font\scap=cmcsc10
\input graphicx

\newcount\eqnumber
\eqnumber=1
\def\neweq{{\rm{(\the\eqnumber)}}\global\advance\eqnumber by 1}
\def\eqdef#1{\eqno\xdef#1{\the\eqnumber}\neweq}
\def\newaeq{{\rm{(\the\eqnumber a)}}\global\advance\eqnumber by 1}
\def\eqdaf#1{\eqno\xdef#1{\the\eqnumber}\newaeq}
\def\eqdisp#1{\xdef#1{\the\eqnumber}\neweq}
\def\eqdasp#1{\xdef#1{\the\eqnumber}\newaeq}

\newcount\refnumber
\refnumber=1
\def\newref{{\the\refnumber}\global\advance\refnumber by 1}
\def\refdef#1{{\xdef#1{\the\refnumber}}\newref}

\newlinechar=`\^^J

\centerline{\bf Confinement strategies in a simple SIR model}
\bigskip
{\scap G. Nakamura},  {\scap B. Grammaticos} and {\scap M. Badoual}

{\sl Universit\'e Paris-Saclay, CNRS/IN2P3, IJCLab, 91405 Orsay, France } 
and {\sl  Universit\'e de Paris, IJCLab, 91405 Orsay France}
\bigskip
{\sl Abstract}
\smallskip
We propose a simple SIR model in order to investigate the impact of various confinement strategies on a most virulent epidemic. Our approach is motivated by the current COVID-19 pandemic. The main hypothesis is the existence of two populations of susceptible persons, one which obeys confinement and for which the infection rate does not exceed 1, and a population which, being non confined for various imperatives, can be substantially more infective. The model, initially formulated as a differential system, is discretised following a specific procedure, the discrete system serving as an integrator for the differential one. Our model is calibrated so as to correspond to what is observed in the COVID-19 epidemic. 

Several conclusions can be reached, despite the very simple structure of our model. First, it is not possible to pinpoint the genesis of the epidemic by just analysing data from when the epidemic is in full swing. It may well turn out that the epidemic has reached a sizeable part of the world months before it became noticeable. Concerning the confinement scenarios, a universal feature of all our simulations is that relaxing the lockdown constraints leads to a rekindling of the epidemic. Thus we sought the conditions for the second epidemic peak to be lower than the first one. This is possible in all the scenarios considered (abrupt, progressive or stepwise exit) but typically a progressive exit can start earlier than an abrupt one. However, by the time the progressive exit is complete, the overall confinement times are not too different. From our results, the most promising strategy is that of a stepwise exit. And in fact its implementation could be quite feasible, with the major part of the population (minus the fragile groups) exiting simultaneously but obeying rigorous distancing constraints. 

\bigskip
MSC2010: 34A34, 37M05, 37N25, 39A30, 92D30
\smallskip
Keywords: epidemics, modelling, SIR model, lockdown

\bigskip
1. {\scap Introduction}
\medskip
Modelling can be a most valuable tool in the investigation of phenomena involving biological systems. This is all the more true when the phenomena in question cannot be reproduced in the laboratory, for practical or ethical reasons. Prominent among the aforementioned are epidemics. Considered as a sub-branch of population dynamics, epidemics lent themselves perfectly to mathematical modelling. The history of such approaches goes back to the 18th century and D. Bernoulli who, in 1760, presented a work [\refdef\danbern] on the mortality rate due to smallpox and the advantages presented  by vaccination. Mathematical tools for the study of epidemics have been developed decades ago and are being constantly refined since. 

The current pandemic sweeping the globe has rekindled the interest in epidemics. Modelling is one of the approaches used in order to identify the best strategy for the mitigation of the devastating effects. The present paper is an attempt at answering, in a modelling framework, the fundamental question of the how to implement efficiently a population confinement. Several works have recently addressed the same question from various perspectives. Four recent papers are the product of the Centre for the Mathematical Modelling of Infectious Diseases (COVID-19 Working Group) of the London School of Hygiene and Tropical Medicine. A stochastic transmission model was used first to estimate the daily reproduction number before and after confinement measures, from data obtained from the Wuhan province. Based on this estimate, the authors of [\refdef\kuchar] conclude that even a small number of infections introduced in a previously unaffected area may lead to an epidemic outbreak. In the second paper [\refdef\hellew], based on the same model, the authors conclude that what plays a major role in determining whether an outbreak is controllable is the delay between the onset of symptoms and the introduction of social distancing measures based on contact tracing. However the necessity of an efficient contact tracing increases with the value of the daily reproduction rate making its practical implementation particularly arduous. Quantifying contacts was the object of the third paper [\refdef\klepac] of the team based on a large-scale data collection concerning the possible (albeit self-reported) contacts of a population of 36000 volunteers. In the fourth paper [\refdef\prem] the authors identify the role of the moment for the relaxation of the physical distancing measures claiming that the benefit of a somewhat delayed relaxation is the reduction of the height of the epidemic peak. Their projections show that a premature lifting of the measures may lead to a secondary peak. A team of Imperial College led by N. Ferguson, examined the two strategies of suppression versus mitigation. While the latter aims at just slowing down the epidemic so as not to overwhelm the healthcare system, the former's objective is to reduce the number of infected to a very low level and maintain it there permanently.  Mitigation is  a more palatable solution (although the authors consider a five month duration of the measures) but it may still lead to a staggering number of deaths and an epidemic resurgence when the interventions are relaxed. The studies [\refdef\imper]  of the Imperial College team show that the timely adoption of measures (be they mitigation or suppression) are of the utmost importance  as far as the consequences of the epidemic are concerned. Any unwarranted delay may result in a huge number of additional infections and deaths.
The team of Stanford under E. Mordecai constructed an interactive model [\refdef\mordec] which simulates the effects of social distancing. Just as in the previous studies a resurgence of the epidemic is predicted if the controls are lifted too quickly. They also model what they call  the `light-switch' approach where confinement is turned on and off in response to the seriousness of the epidemic resurgences. A team of the University of Oxford [\refdef\gupta] applied modelling techniques to the data collected in Italy and the UK. Their conclusion is that the epidemic has started at least a month before the first reported death and would have an approximate duration of 2-3 months, in the absence of any intervention. On the positive side, their analysis seems to indicate an accumulation of significant levels of herd immunity. 

The particularity of our modelling approach resides in the fact that we are using the simplest possible model for this, namely, an adaptation of the venerable SIR. The SIR model was proposed by Kermack and McKendrick [\refdef\kermac], inspired from the works of Ross [\refdef\ross]. The basic model consists in a population split into three groups: `healthy' individuals that are susceptible to infection ($S$), `infected' individuals who can transmit the disease ($I$), and the `removed' ($R$) who either died from the disease or, having recovered, are immune to it. The differential model built on the assumption of a fully mixed population has the form
$${dS\over dt}=-aSI$$
$${dI\over dt}=aSI-\lambda I\eqdef\zena$$
$${dR\over dt}=\lambda I$$ 
with $a$ being the infection rate and $\lambda$ the removal rate of the infected individuals. The main assumption here is that the number of infected individuals increases at a rate proportional to the number of both infected and healthy. The model equations (\zena) are such that the total population $S+I+R$ is constant. By introducing the appropriate scaling of the time $t$ it is possible to set the parameter $\lambda$ to 1. In this case the value of $a$ defines what is called the basic reproduction number i.e. the expected number of infections in the susceptible population resulting form a single infection. 

There exist a slew of versions of the SIR model [\refdef\extend], introducing more subsets of the population and/or different interactions among them. Hethcote [\refdef\hethcot] presents a pedagogical review of the matter. Of particular interest to our approach is the fact that the SIR model can be modified so as to model vaccination. As explained clearly in [\refdef\bauch], a vaccination policy presents a challenge. If a large proportion of the population is already immune then the risk, even a small one, associated with vaccination will be perceived as more significant than the risk from infection. This resonates with the perceived risks of social distancing measures, where a lockdown although offering protection from infection also entails economical and psychological downsides.  

In what follows we shall introduce our version of the SIR model, particularly tailored for the investigation of the effect of lockdown and its eventual relaxation. We shall derive our special discrete version of the model and will use it to perform the numerical simulations. Based on the latter we shall discuss the effect of lockdown and its relaxations scenarios. 
\bigskip
{\scap 2. The model}
\medskip
The classical SIR model introduces three populations: susceptibles $S$, infective $I$ and recovered $R$. Given the conservation of the total population the equation for the recovered is superfluous. In view of the investigation of confinement strategies we separate the susceptible population into two sub-groups, the confined-susceptibles, $C$, and the unconfined-susceptibles, $U$. The equations become now
$$C'=-aCI$$
$$U'=-bUI\eqdef\zdyo$$
$$I'=aCI+bUI-\lambda I,$$
and by scaling time we put $\lambda=1$. Here $C$ and $U$ are to be understood as percentages of the total population, with $C(0)+U(0)\approx1$ (no `recovered' at the beginning of the epidemic). The condition for the number of infectives to grow is $aC(0)+bU(0)>1$, otherwise the epidemic fizzles out. Practically, since the measures to curb the epidemic are taken once the latter has manifest itself, the initial population is unconfined.  In this case the condition for an epidemic outbreak is simply $b>1$. A fixed point of equation (\zdyo) has necessarily $I_*=0$, but we can have $C_*U_*\ne0$. The fixed point is attractive provided $aC_*+bU_*<1$. Practically this means that the epidemic stops when there are no more infective persons, while a non-infected fraction the population may exist.

Since we are dealing with populations the quantities $C$, $U$ and $I$ are a priori positive. Given the structure of the equations (\zdyo) positivity is guaranteed provided one starts from positive initial conditions [\hethcot]. Taking this into account we shall proceed now to the construction of a discrete version of (\zdyo). Our approach is inspired by the works of Mikens [\refdef\mikens]. His recommendations are that `nonlinear terms must be, in general, replaced by nonlocal discrete representations' and `a property that holds for the differential equation should also be present in the discrete model'. Our rule of thumb in this is [\refdef\handy] that `since all quantities are positive, no minus sign should appear anywhere'. We introduce thus a forward difference of the time derivative, with time step $\delta$, and, what is more important, the staggering below
$${C_{n+1}-C_n\over\delta}=-aI_nC_{n+1}$$
$${U_{n+1}-U_n\over\delta}=-bI_nU_{n+1}\eqdef\ztri$$
$${I_{n+1}-I_n\over\delta}=aI_nC_{n+1}+bI_nU_{n+1}-I_{n+1}.$$
Solving for the points at $(n+1)$ we obtain
$$C_{n+1}={C_n\over1+a\delta I_n}$$
$$U_{n+1}={U_n\over1+b\delta I_n}\eqdef\ztes$$
$$I_{n+1}={I_n(1+a\delta C_{n+1}+b\delta U_{n+1})\over1+\delta}.$$
This system has the very same properties as the differential one. In what follows we shall use it in order to integrate the differential system (\zdyo). Moreover (\ztes) is a very robust scheme which gives a realistic answer for {\sl any} time step, even a very large one. Let illustrate this with an example. We perform a simulation taking a time step $\delta=10^{-3}$, without distinguishing between confined and unconfined populations, $a=b=2.$ and plot the infective fraction of the population, $I$, as well as the fraction of infected, $1-(C+U)$ as a function of time.
\bigskip
\centerline{\includegraphics[width=10cm,keepaspectratio]{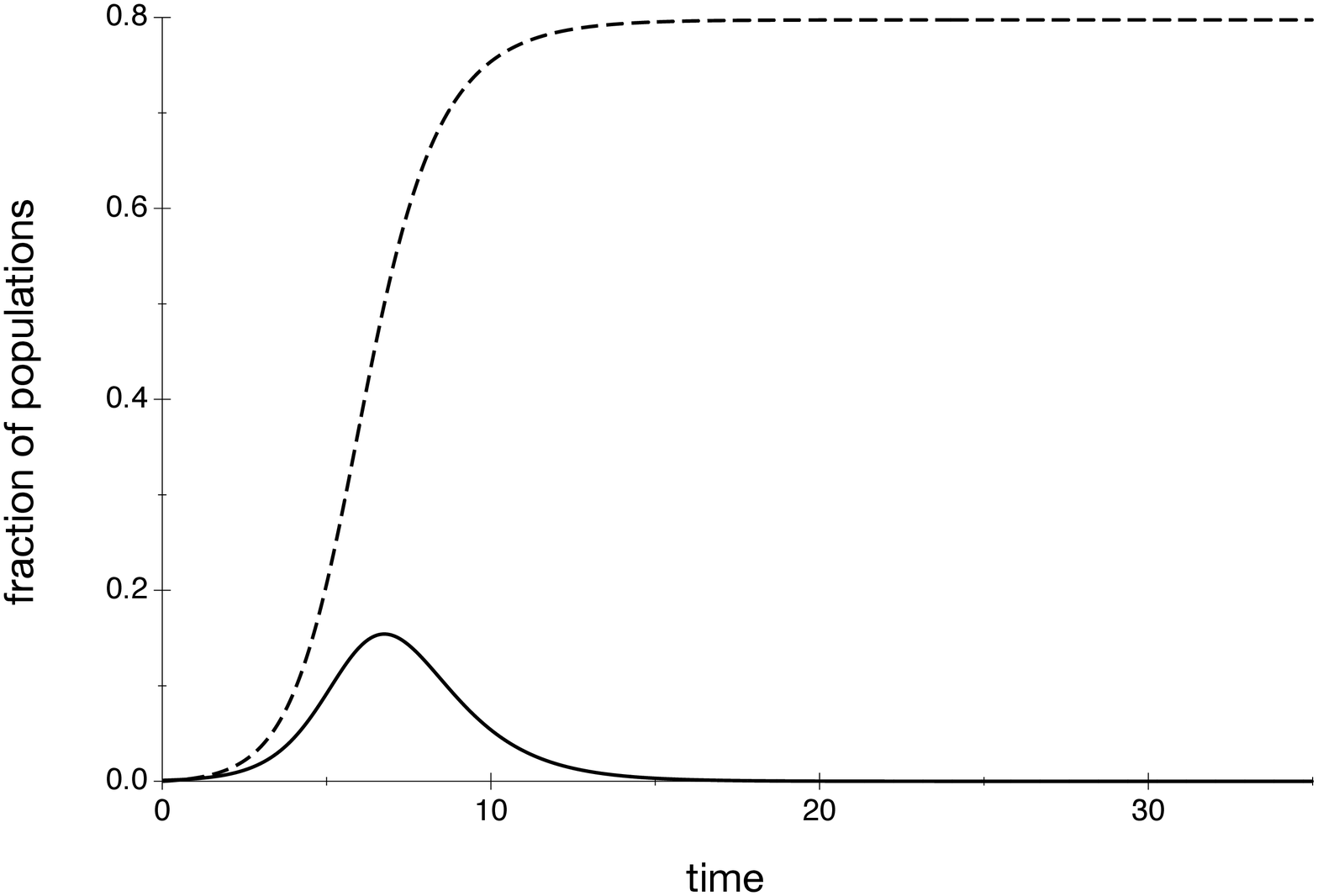}}
\smallskip{\bf Figure 1}. {\it The infective fraction (full line) and the infected fraction (dashed line) of the population as a function of time.}
\medskip
Next we consider a case where the time step can become large. We introduce $J=\delta I$ and rewrite equation (\ztes) after taking the limit $\delta\to\infty$. We obtain now the system
$$C_{n+1}={C_n\over1+a J_n}$$
$$U_{n+1}={U_n\over1+b J_n}\eqdef\zpen$$
$$J_{n+1}=J_n(a C_{n+1}+b U_{n+1}).$$
We iterate this system with the same conditions as for Figure 1, and obtain the result shown in Figure 2. 
\bigskip
\centerline{\includegraphics[width=10cm,keepaspectratio]{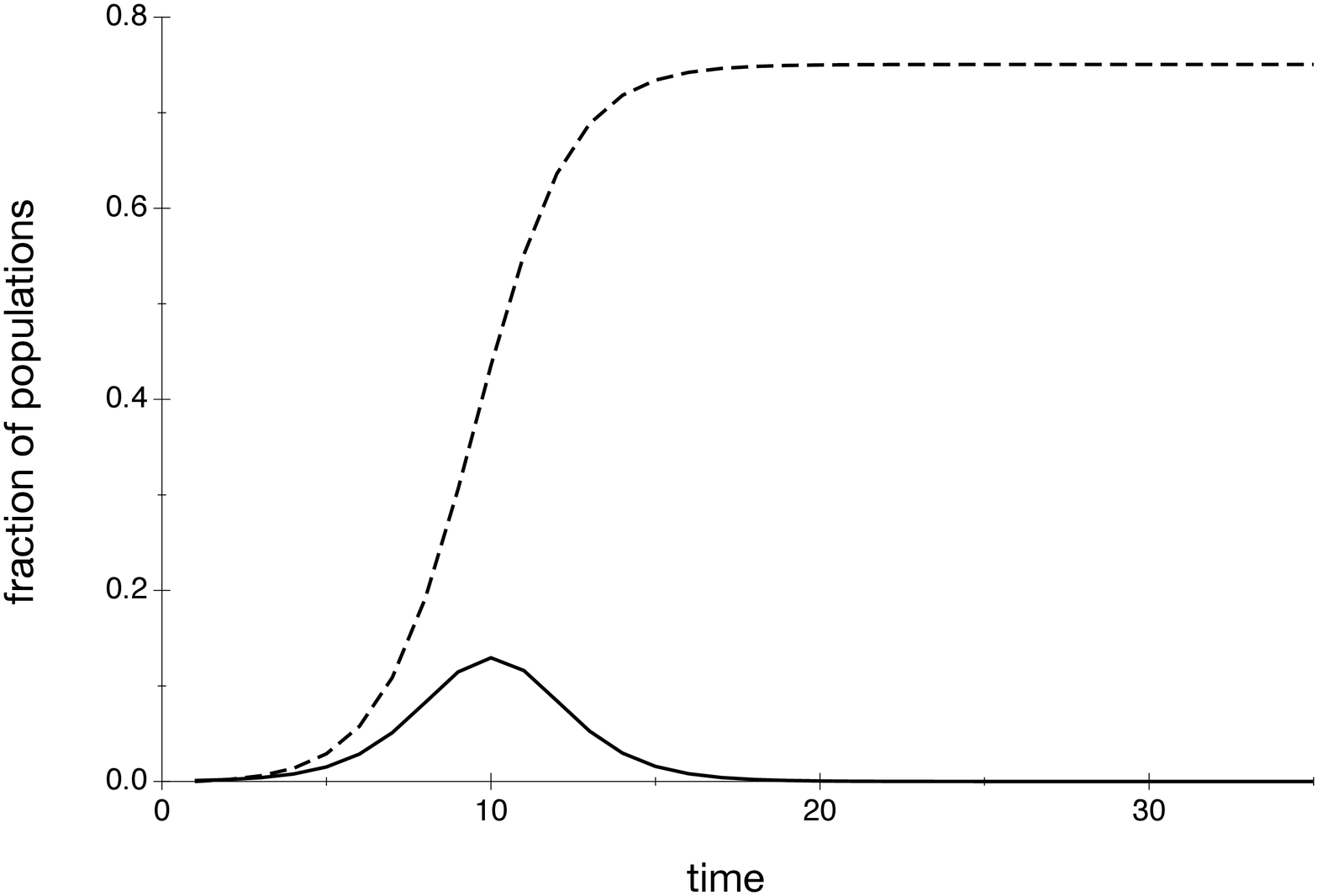}}
\smallskip{\bf Figure 2}. {\it Same as for Figure 1, but obtained at the limit of an infinite time step.}
\medskip
The result is striking. While some details, like the value and position of the peak or the asymptotic value of infected population, may differ slightly, the overall behaviour of the results at infinite time step is very close to that of the small time step used in order to simulate the continuous system. This is the definite advantage of tailored integrators as compared to ready-made, black-box, routines.

Before proceeding to the examination of the various confinement strategies it is interesting to apply our model to a simple situation, that of hunting for the `patient zero`. We have performed a simulation of (\zdyo) assuming a wholly unconfined population, i.e. $C(0)=0, U(0)=1$, with an infection rate $b=2$. The results of the simulation with a `reasonable' percentage of infectives $I(0)=10^{-3}$ are those shown in Figure 1. Next we performed a simulation assuming that one person on earth was infected, the patient zero. The corresponding initial condition is $I(0)\approx10^{-10}$. The result of the simulation is shown in Figure 3.
\bigskip
\centerline{\includegraphics[width=10cm,keepaspectratio]{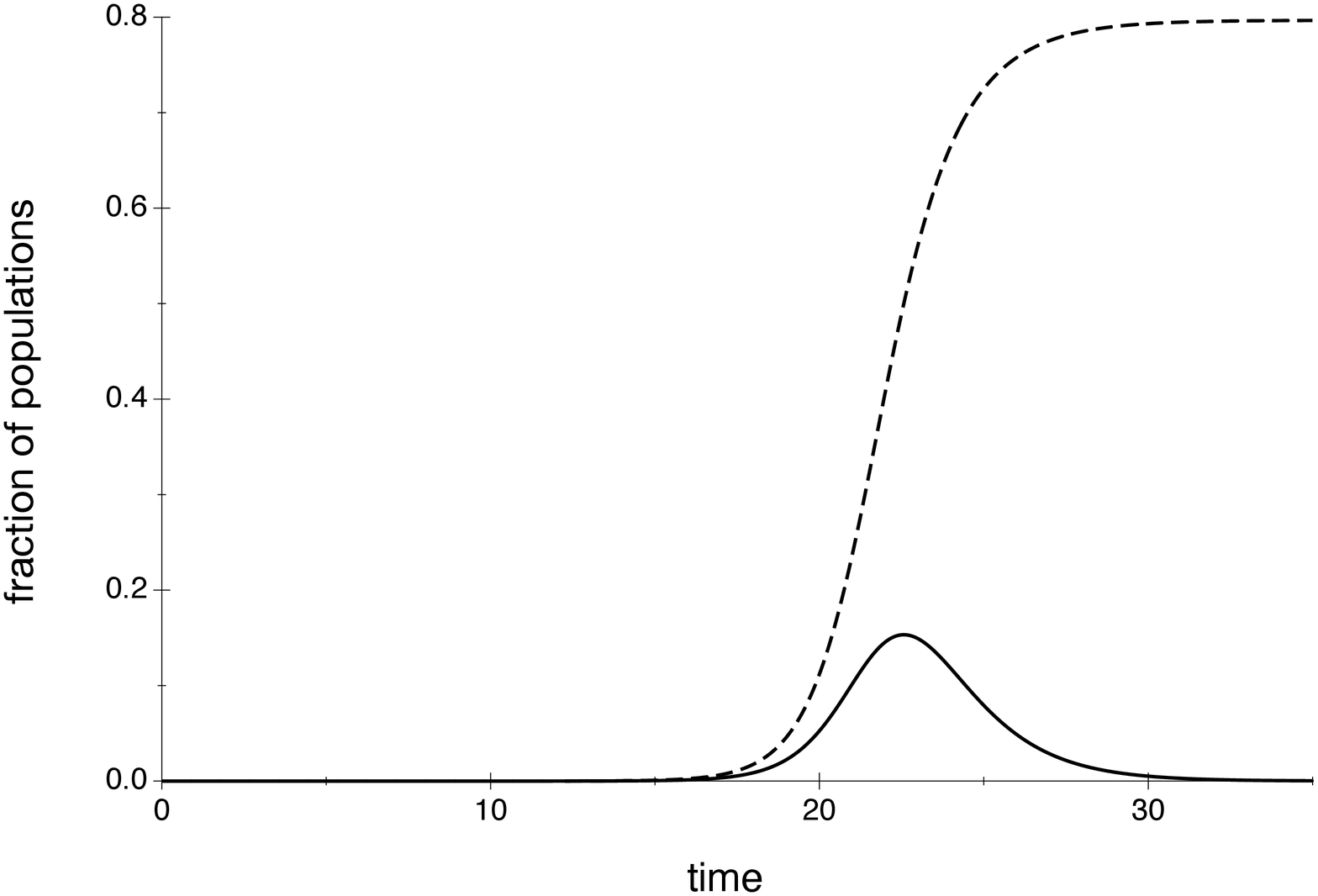}}
\smallskip{\bf Figure 3}. {\it The infective fraction (full line) and the infected fraction (dashed line) of the population as a function of time starting from initial conditions corresponding to the `patient zero'.}
\medskip
The comparison is quite interesting: the profile of the epidemic, i.e. the evolution of the infectives over time, is the same with both initial conditions, just shifted in time. Thus by following the evolution of the epidemic one cannot go back in time and guess when the first infection took place. This has to be assessed by other methods which lie beyond the modelling approach. Put in a different way, by observing the evolution of the epidemic starting at a given time one cannot be sure where the genesis of the epidemic is situated in time.
\bigskip
{\scap 3. Model predictions}
\medskip

In what follows we shall apply our model to situations where a population is totally or partially confined and where the lockdown is lifted progressively or abruptly. In order for our simulations to be as realistic as possible it is important that we calibrate our model, introduce the proper time scale, choose the proper parameters and initial conditions, and, finally consider the adequate confinement strategies.
\medskip
{\sl Time scale}
\smallskip
The time scale, given by $1/\lambda$, is set to 5 days. This is a typical time for the duration of infectiousness,  according to Ferguson and collaborators [\imper]. In what follows, time is counted in units of $1/\lambda$.
\medskip
{\sl Parameter values}
\smallskip
The value of the basic reproduction number without confinement $R_0={b/ \lambda}$ is still not known with precision. In [\refdef\li], the $R_0$ was estimated to be around 2.2 and in [\refdef\zhao], $R_0$ ranges from 2.24 to 3.58. 
In what follows (having taken $\lambda=1$) we shall choose $R_0=b=2.6$.
The goal of confinement is to reduce $R_0$ to a value close or even below 1, which is the threshold for an epidemic outbreak [\imper].
The infection rate for strictly confined people $a=a_0$ was set to 1.
\medskip
{\sl Initial conditions}
 \smallskip
 We set the fraction of confined people to 0.7. This value seems reasonable, since  some people go to their workplace every day. These people are more exposed to the virus and belong to the category of unconfined people. In France, for example, it is estimated that around a quarter of active people still go to work.
 
The initial value of the fraction of infected people is set to $10^{-5}$, and the confinement is applied after a time interval, $T_0$. Since we are interested here in the different strategies of ending the confinement, $T_0$ is fixed in all the simulations, $T_0=4$ (or 20 days). 

The time step used in the simulations is $\delta=10^{-3}$.  

We tested our choice of parameters and initial conditions by comparing the evolution of the fraction of infected people to the real evolution of active cases in Italy. The confinement started in Italy on March 8th in the north of the country, so we report here the data of the active cases from 20 days before this date (February 19th 2020), to April 16th, 2020 [\refdef\italy]. We divided these numbers by the population of the four most impacted regions of northern Italy (Lombardia, Piemonte, Emilia Romagna, Veneto), which is around 20 millions of people, so as to get a density. In order to get a good match with the model, we had to correct the density by a factor of 9, which is consistent with the fact that a large fraction of cases are actually unreported or asymptomatic cases [\refdef\lili]. 
 \bigskip
\centerline{\includegraphics[width=10cm,keepaspectratio]{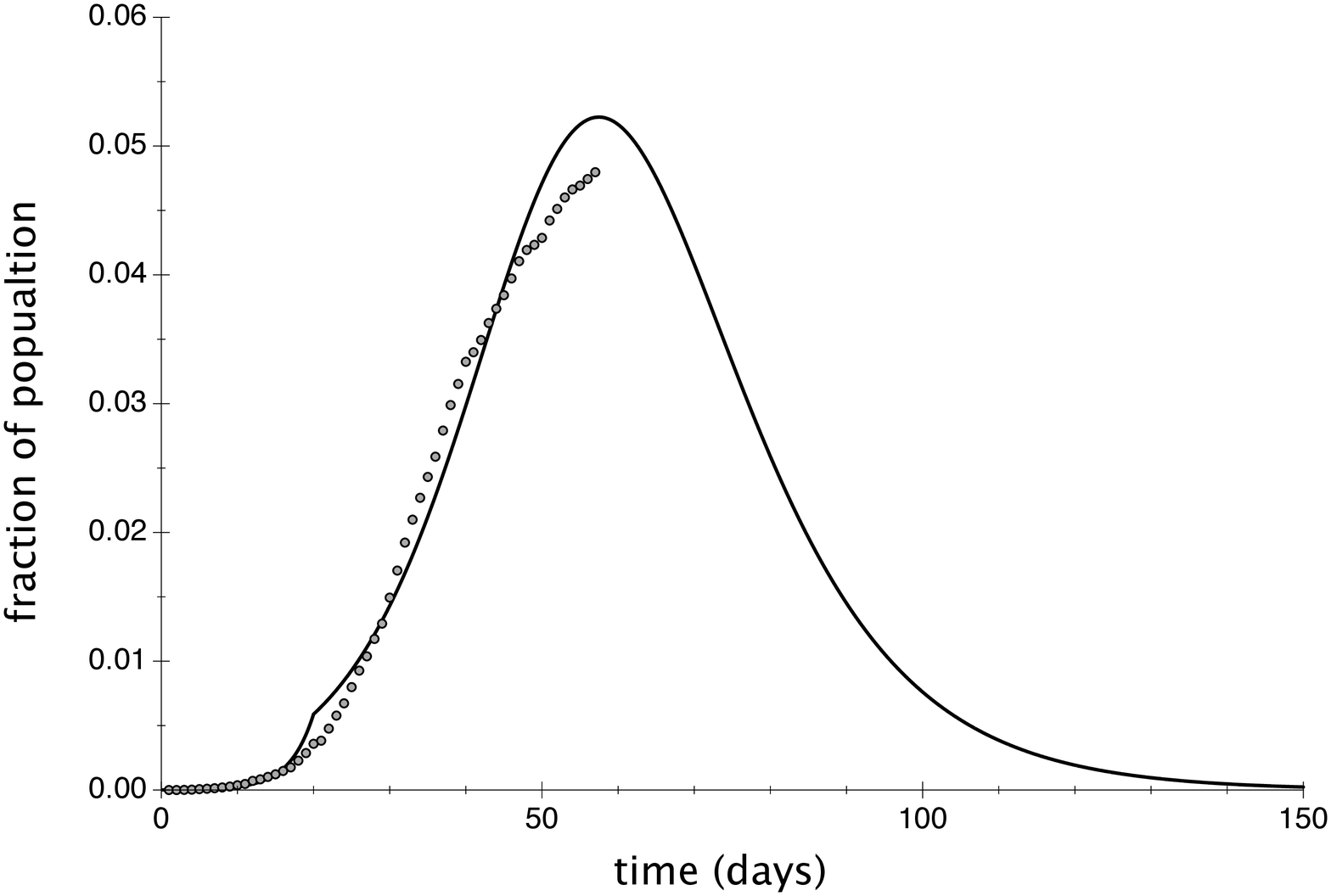}}
\smallskip{\bf Figure 4}. {\it Comparison between the temporal evolution of the fraction of infective population obtained by simulation and of the real number of active cases in Italy, between February 19th 2020 and April 16th, 2020.}
\medskip
The agreement between the real data and our model is satisfactory, we can conclude that our parameters and initial conditions are realistic. 
\medskip
{\sl Different strategies for lifting confinement}
 \smallskip
Once the model is calibrated we are ready to address the question of population confinement. The first simulation is performed so as to compare the two extreme cases, one where there is no lockdown and one where the duration of the lockdown is extended till the practical disappearance of the epidemic. The results are presented in Figure 5. 

\bigskip
\centerline{\includegraphics[width=10cm,keepaspectratio]{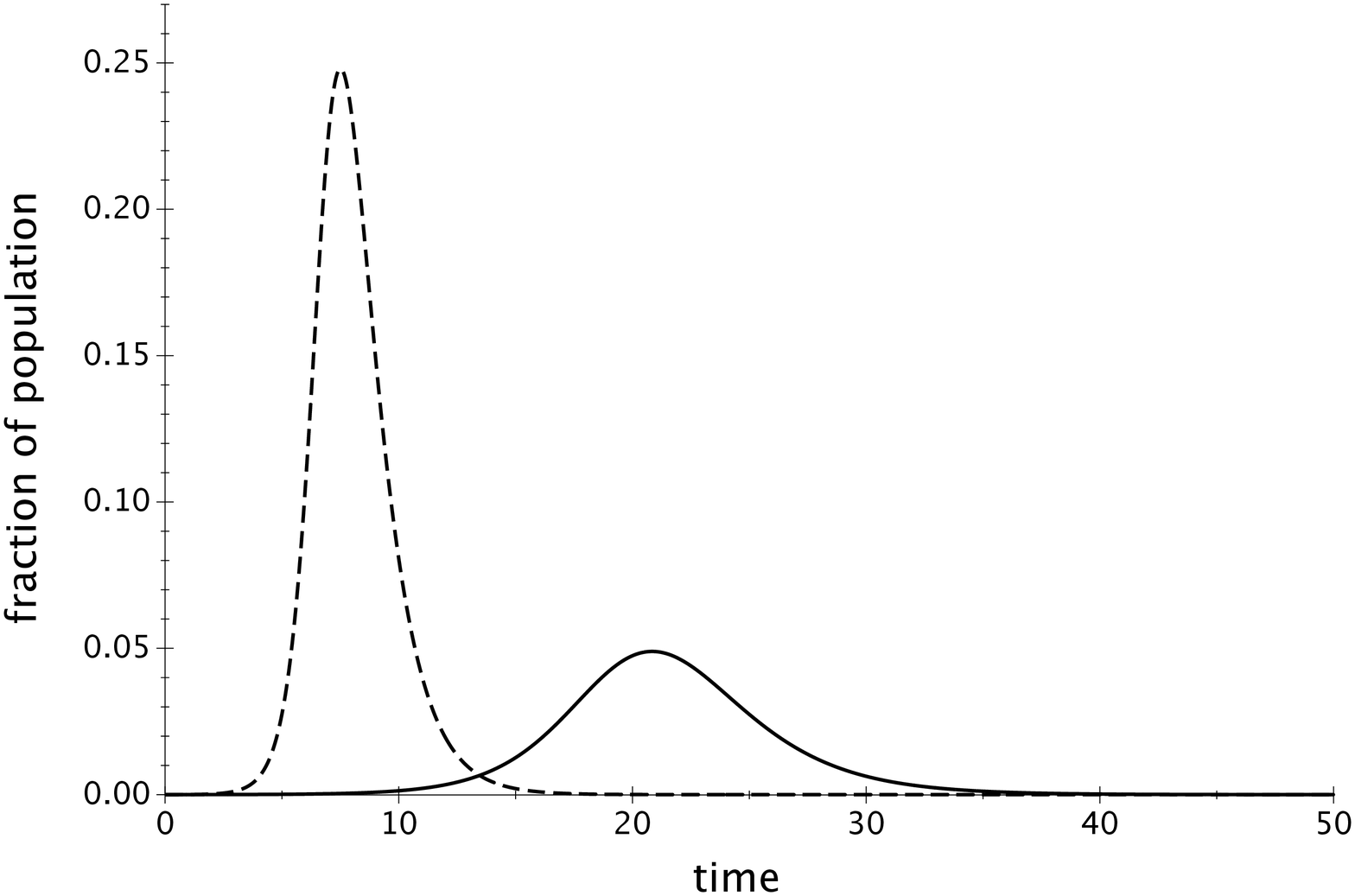}}
\smallskip{\bf Figure 5}. {\it The lines represent the temporal evolution of the infective fraction of population, for the linear exit of confinement (full line) and for the absence of any measure of confinement (dashed line). The confinement starts after $T_0=4$ and lasts for the whole time of the simulation.}

\medskip
They correspond in fact to the curves repeatedly shown by the media. With an absence of confinement the epidemic is of short duration but of high intensity: a sizable part of the population is infected. The positive side of such a strategy is that herd immunity of the population is reached in a short time. The downside is that the healthcare system can easily be overwhelmed by the sheer number of infected people. The prolonged confinement has the advantage of making the number of infected people (hopefully) manageable. On the downside, such a prolonged lockdown may have untenable social and financial repercussions. Thus alternative strategies must be investigated.

The confinement strategies we are going to investigate in what follows have to do with the lifting of lockdown measures. We shall not delve into the initial application of lockdown. The only visible strategy adopted universally is to enforce such a measure when it clear that the epidemic cannot be ignored any more. However exiting the confinement is particularly delicate and thus we are going to focus our analysis on this.

Three different scenarios will be examined. They correspond to an abrupt, instantaneous exit, to a progressive, continuous, one and to a progressive exit by steps.  
We model the effect of confinement by changing the infection rate of the confined population $a$ during time. Before the onset of confinement, this population has the same infection rate as the unconfined one, so $a(t)=b$.  During strict confinement, $a(t)=a_0$ with $a_0<b$.
This period of strict confinement lasts for a period of time $T_1$. After this time interval, the lockdown starts to be lifted and a period of lighter confinement starts. The duration of this period of time is defined as $T_2$. Different strategies of confinement lifting can be tested: the abrupt ending of confinement, where $a(t)$ changes instantaneously from $a_0$ to $b$ and $T_2=0$; the progressive ending of confinement, where $a(t)$ increases linearly during $T_2$ and the step ending, where $a(t)$ takes an intermediate value $a_1$, between $a_0$ and $b$ during $T_2$. The function $a$ and the time intervals $T_0$, $T_1$ and $T_2$ are defined on Figure 6.

In our simulations, the parameters $b, a_0$ and $T_0$ are fixed: $b=2.6$, $a_0=1$ and $T_0=4$. The parameters $a_1$, $T_1$ and $T_2$ can vary.

\bigskip
\centerline{\includegraphics[width=10cm,keepaspectratio]{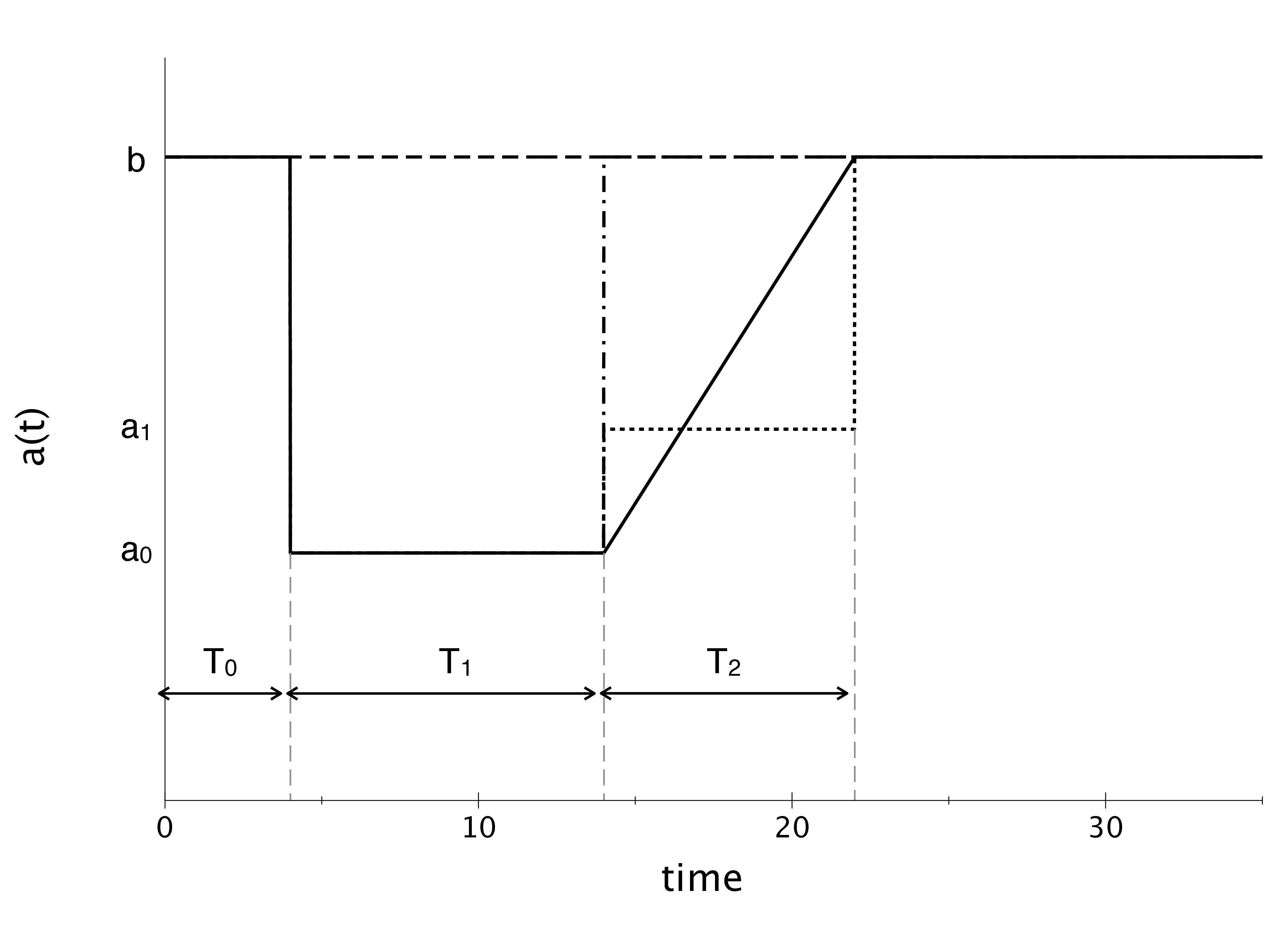}}
\smallskip{\bf Figure 6}. {\it  The  lines represent the temporal evolution of the infection rate of confined people, $a(t)$, for the different strategies of exit of confinement. The full line is for the linear exit, the dashed-dotted one for the abrupt exit, the dotted one for the stepwise exit. The dashed line is for the case without any confinement. The time intervals of no confinement $T_0$, strict confinement, $T_1$, and of lighter confinement during the exit $T_2$ are indicated.}
\medskip

Figure 7 presents the results of a simulation where we compare an abrupt confinement exit to a continuous one. 
This is a most interesting result, since it shows that misjudging the appropriate moment for the exit may result in a second wave of the epidemic worse than the first one. 
\bigskip
\centerline{\includegraphics[width=10cm,keepaspectratio]{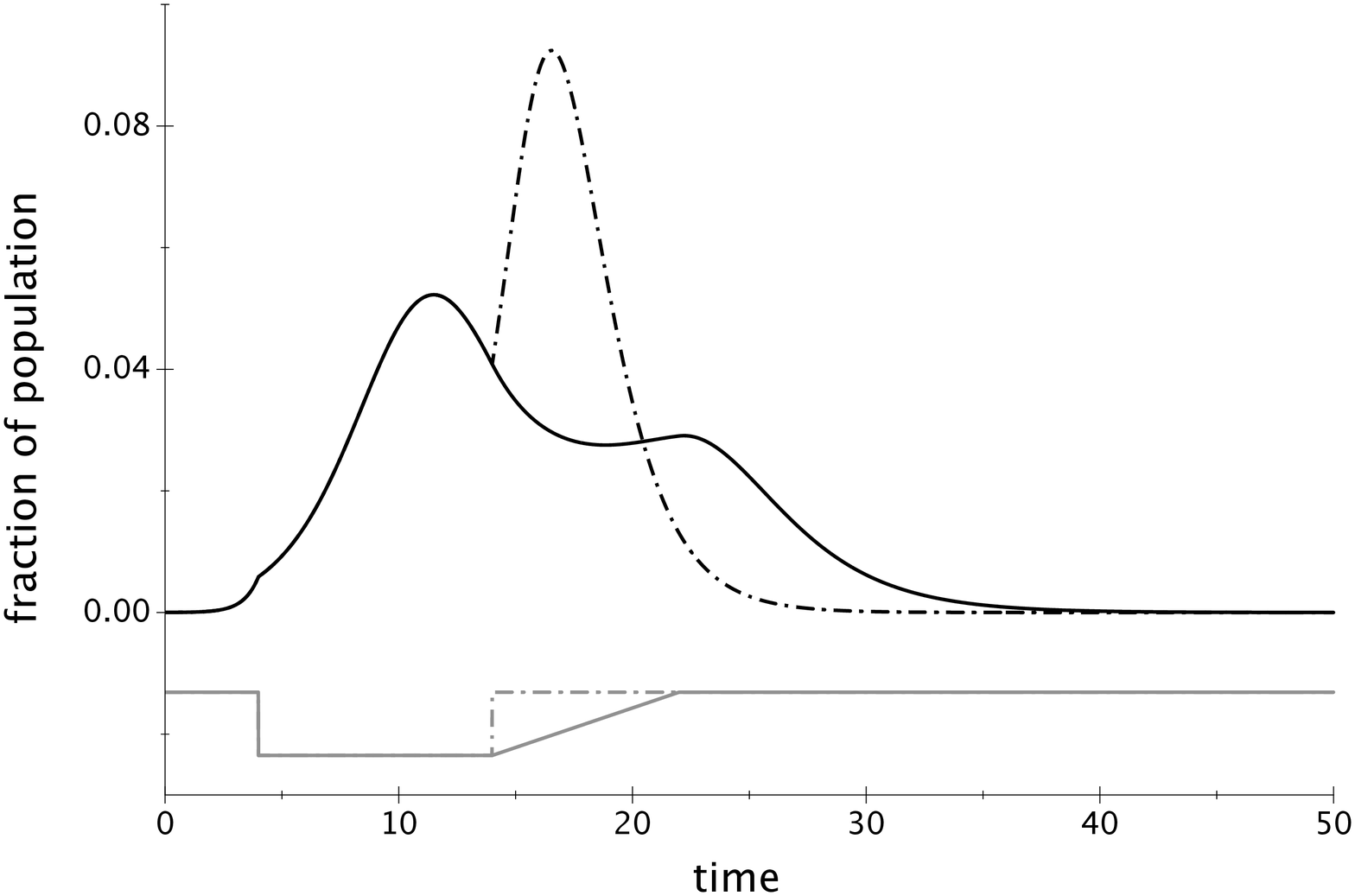}}
\smallskip{\bf Figure 7}. {\it The black lines represent the temporal evolution of the infective fraction of population, for the linear and abrupt exit of confinement. The grey lines in the lower part of the figure represent the temporal evolution of the infection rate of confined people, $a(t)$ (with $T_1=10$ and $T_2=8$). The full lines (black and grey) are for the linear exit of confinement and the dashed-dotted ones (black and grey) are for the abrupt exit of confinement. }
\medskip

On the other hand choosing the same moment for the relaxation of the lockdown measures but applying a progressive scenario keeps everything manageable despite a slight recurrence of the increase of the number of infected people. A second peak of infection does appear but it is lower than the first one.

A stepwise exit from confinement is shown in Figure 8, in comparison with the progressive (linear profile) exit. 
\bigskip
\centerline{\includegraphics[width=10cm,keepaspectratio]{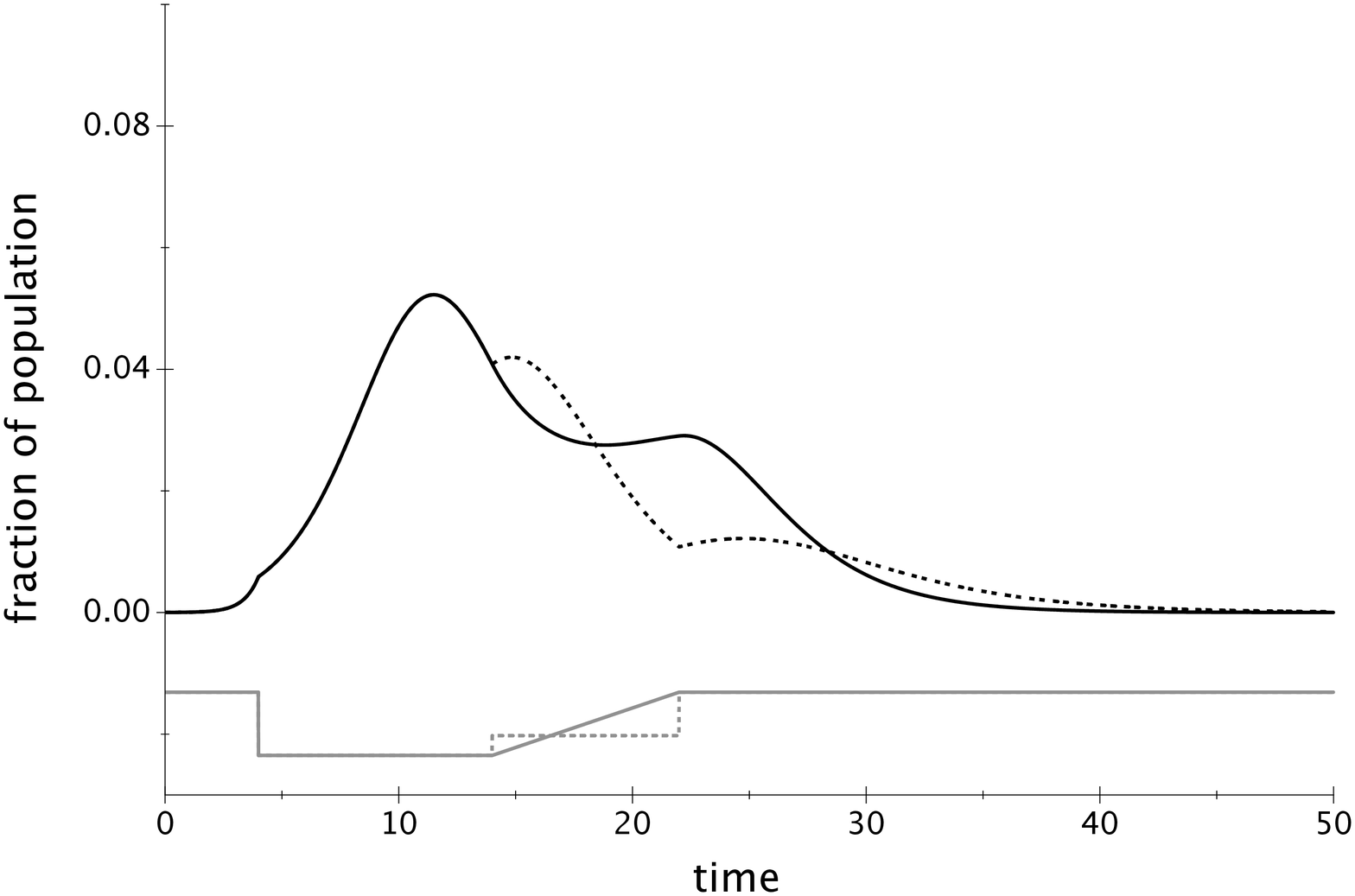}}
\smallskip{\bf Figure 8}. {\it The black lines represent the temporal evolution of the infective fraction of population, for the linear and stepwise exit of confinement. The grey lines  in the lower part of the figure represent the temporal evolution of the infection rate of confined people, $a(t)$ (with $a_1=1.5, T_1=10$ and $T_2=8$). The full lines (black and grey) are for the linear exit of confinement and the dotted ones (black and grey) are for the stepwise exit of confinement.}
\medskip
The stepwise strategy has the particularity to lead to a non-negligible increase in the number of infected people right after the first, partial, exit, but, on the other hand, it presents the advantage that by the time the exit is total the level of the epidemic can be considered as wholly manageable. 
In the case of the stepwise strategy, there are actually three peaks in the infection curve, each one corresponding to a discontinuity of function $a$. However, the third peak is always much smaller than the first two. Thus, in what follows, we will compare only the first peak to the second one.

In Figure 9 we show the variation of the (fractions of the) populations of the confined and unconfined susceptible as a function of time for two exit scenarios. As expected, an abrupt exit has as a consequence a sharp decline in the population of those who were previously confined. Being now in contact with the infectives they can become infected, which explains the sharp increase observed in Figure 7. A progressive exit, although also leading to increased infections, has a much smaller effect and a definitely smoother variation of the susceptible population.
\bigskip
\centerline{\includegraphics[width=10cm,keepaspectratio]{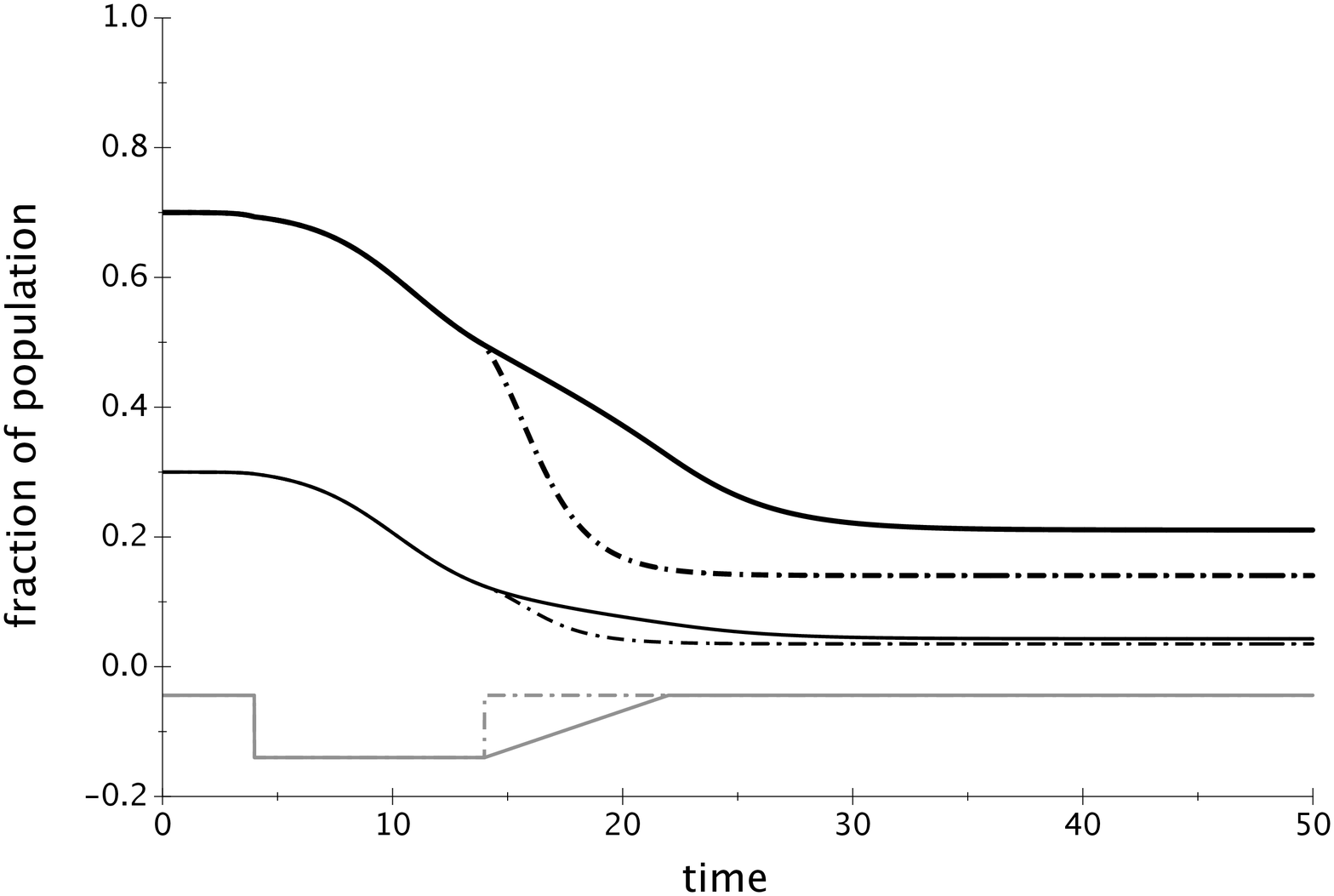}}
\smallskip{\bf Figure 9}. {\it The black lines represent the temporal evolution of the confined and unconfined fraction of population, for abrupt and linear exit of confinement. The thick lines (full and dashed-dotted) correspond to the confined fraction of population, and the thin lines to the unconfined one. The grey lines  in the lower part of the figure represent the temporal evolution of the infection rate of confined people, $a(t)$ (with $T_1=10$ and $T_2=8$). The full lines (thick black, thin black and grey) are for the linear exit of confinement and the dashed-dotted ones (thick black, thin black and grey) are for the abrupt exit of confinement.}
\medskip
Figure 10 summarises the behaviour of the total population of infected people for the four scenarios considered here, no-confinement, confinement with abrupt exit, progressive exit and stepwise exit. As expected the number of globally infected people is larger in a no-confinement scenario. This is the one offering the most efficient `herd-immunity' but the price to pay is a high number of infections concentrated in time. The abrupt exit scenario has a relatively smaller number of  infected people, but as we saw it presents the danger of a second epidemic wave being worse than the first. The progressive strategies are the ones leading to the smallest number of infected people. While this has the advantage of keeping the epidemic manageable all along it is not optimal if what was sought to begin with was an as large as possible immunisation of the population. 
\bigskip
\centerline{\includegraphics[width=10cm,keepaspectratio]{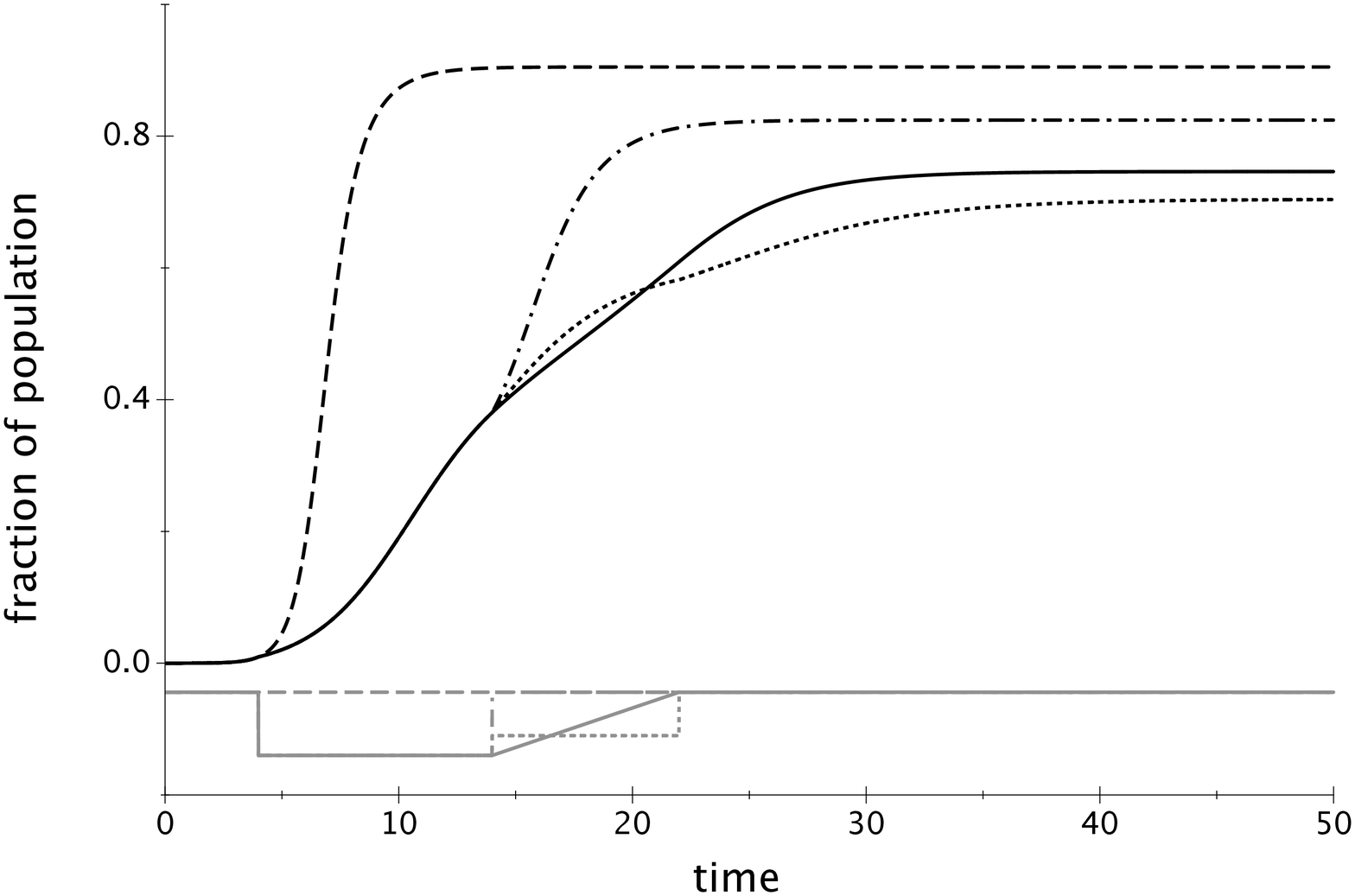}}
\smallskip{\bf Figure 10}. {\it The black lines represent the temporal evolution of the fraction of population that is or has been infected $(1-C-U)$, for the different lockdown exit strategies. The grey lines in the lower part of the figure represent the temporal evolution of the infection rate of confined people, $a(t)$ (with $a_1=1.5, T_1=10$ and $T_2=8$). The dashed lines (black or grey) correspond to the absence of any measure of confinement, the dashed-dotted lines (black or grey) correspond to the abrupt exit of confinement, the dotted lines (black or grey) to the stepwise exit and the full lines (black or grey) to the linear exit.}
\medskip
The effects of the lockdown duration as well as that of the exit strategy are better represented in a figure summarising a broad collection of values. Thus in Figure 11 we show the ratio of the second to the first epidemic peak, i.e. the one reached after the exit from lockdown to the one obtained during the confinement, as a function of the duration of the strict confinement, $T_1$. The curves labelled as (a), (b) etc. correspond to increasing progressive durations of the lockdown exit $T_2$, while the dashed curve is the one obtained for an abrupt exit. Clearly the latter would requite a more prolonged confinement period for the second epidemic peak not to be higher than the first. 
One can notice that since we study the different scenarios of ending confinement, the height of the first peak, reached during confinement, before starting to lift the constrains, is the same for all the values of the studied variables. 
\bigskip
\centerline{\includegraphics[width=10cm,keepaspectratio]{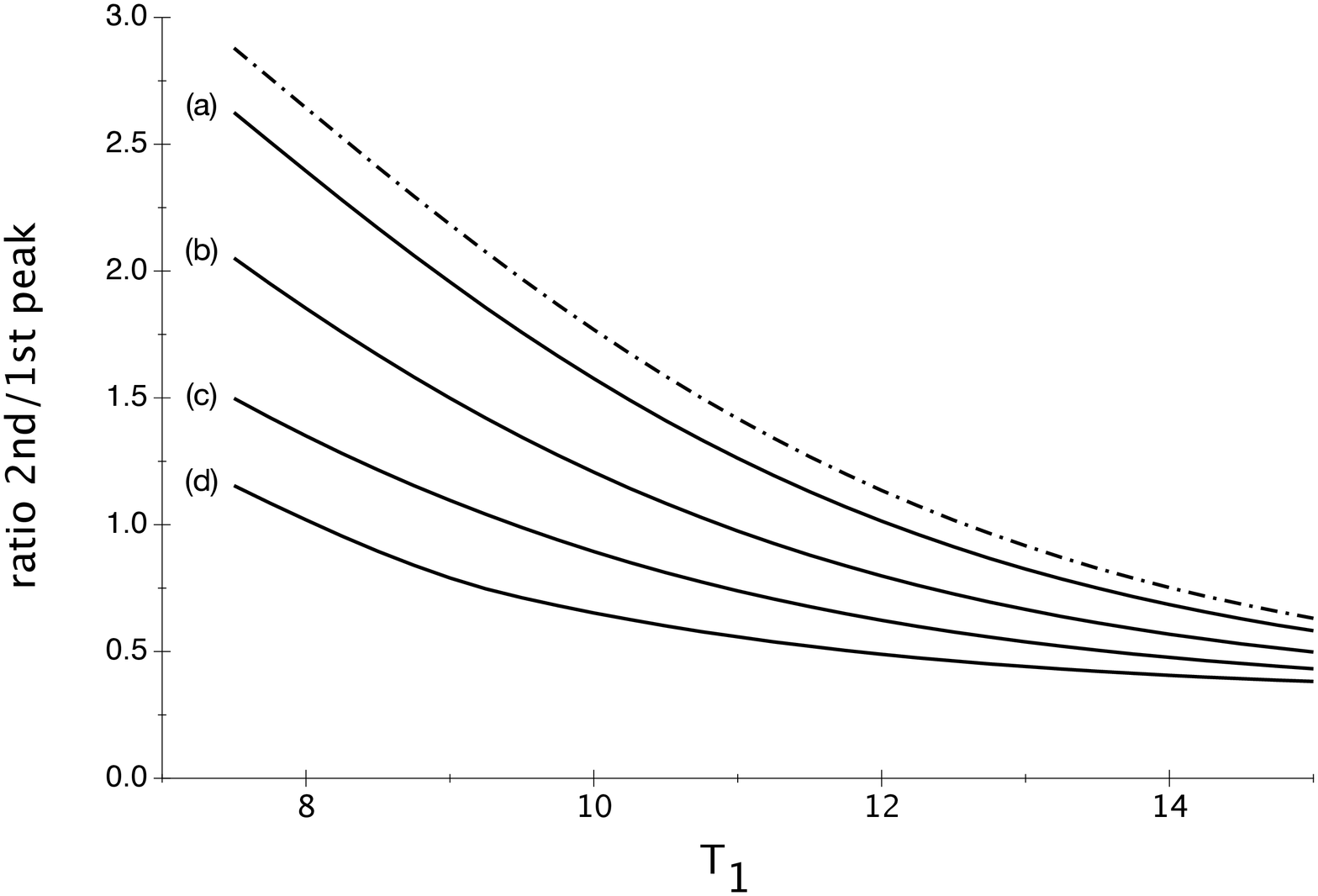}}
\smallskip{\bf Figure 11}. {\it The curve represent the variation of the ratio of the second to the first peak of the infection curve as a function of the duration of the strict lockdown $T_1$. Curve (a) corresponds to the duration of the progressive exit (or lighter confinement period) $T_2=1$, curve (b) to $T_2=3$, curve (c) to $T_2=5$ and curve (d) to  $T_2=7$. The dashed curve corresponds to the abrupt exit (equivalent to $T_2=0$). For all the curves, $T_0=4$. }
\medskip
Figure 12 concerns the stepwise strategy. In this case, the infection rate of confined people takes an intermediate value before going back to unconfined value. In Figure 12, we show the ratio of the second to the first peak in the infection curve, as a function of the intermediate value of the infection rate of confined people. 
\bigskip
\centerline{\includegraphics[width=10cm,keepaspectratio]{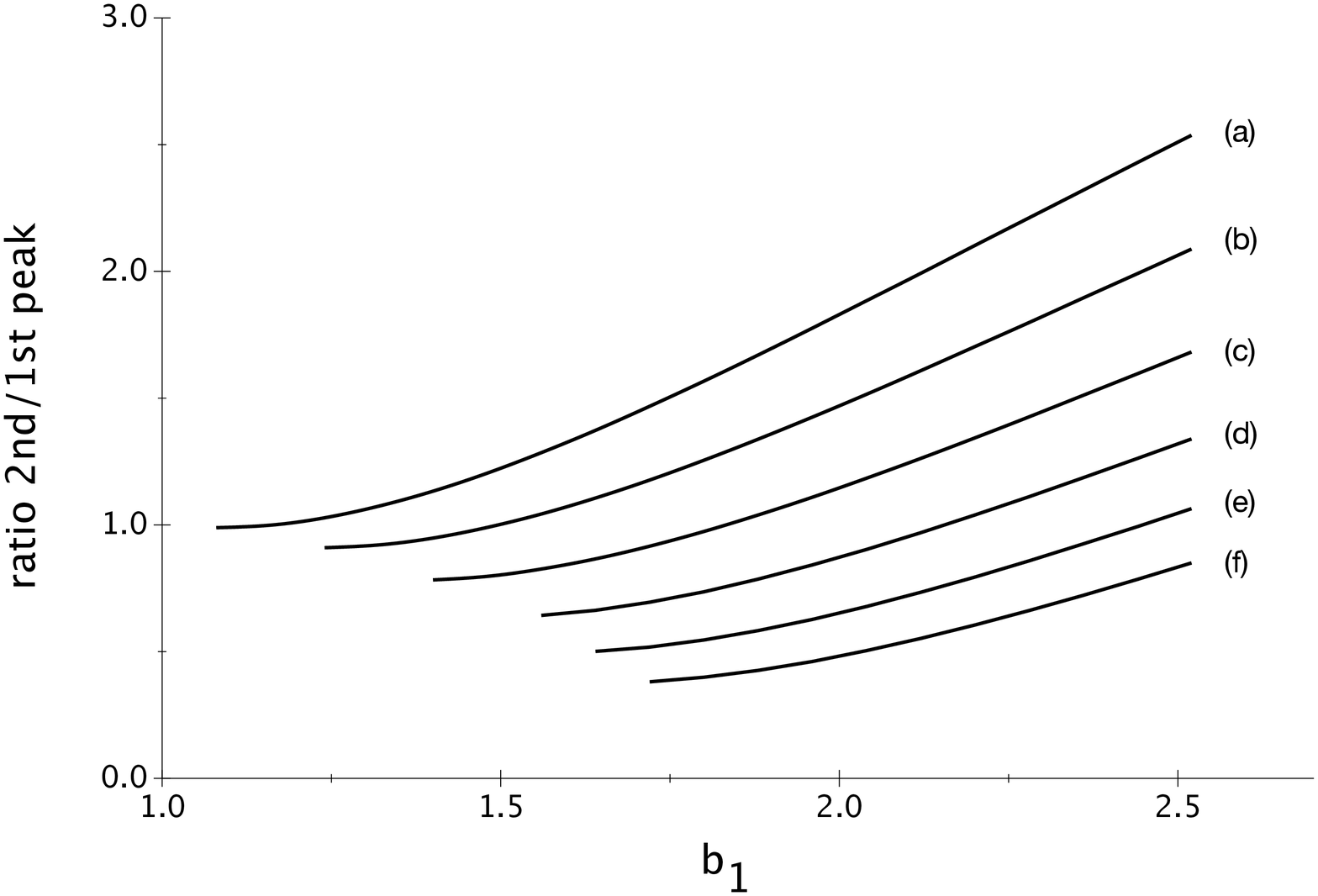}}
\smallskip{\bf Figure 12}. {\it The curves represents the variation of the ratio of the second to the first peak of the infection curve as a function of the intermediate value of the infection rate for confined people, $a_1$. Curve (a) corresponds to the duration of confinement $T_1=8$, curve (b) to $T_1=9$, curve (c) to $T_1=10$, curve (d) to  $T_1=11$, curve (e) to  $T_1=12$,  curve (f) to  $T_1=13$. For all the curves, $T_0=4$ and $T_2=8$. }
\medskip
The different curves (a), (b), (c), (d), (e) and (f) correspond to different values of the confinement duration, $T_1$. The duration of the intermediate state of confinement, $T_2$, where $a=a_1$, is constant for all the curves of the figure and is equal to $8$ (since the height of the third peak is not studied, this value is not important, as long as the duration is long enough so that the third peak does not interfere with the second one). 
\bigskip
{\scap 4. Conclusions}
\medskip
In this paper we have introduced a variant of the SIR model in order to study, in the simplest possible way, without unnecessary assumptions, the effect of population confinement on the evolution of an epidemic. The motivation for this is, obviously, the current COVID-19 epidemic, which in a matter of a few months has evolved into a pandemic. The gravity of the situation, has made mandatory the introduction, in most countries, of lockdown measures aiming at containing the epidemic, so as not to overwhelm fragile healthcare systems. Our model is a classical SIR one, where the population of the susceptibles is spilt into two subgroups, one  which is essentially confined and one, which, due to specific duties, cannot follow strictly the social distancing directives. The infectiveness of the second population, the presence of which is crucial for the functioning of society, in sufficient in order to spread the epidemic. 

Our model was formulated as a system of three coupled differential equations. We discussed the conditions for an epidemic to evolve, the existence of fixed points as well as their stability. A discrete form of the equations was introduced which was subsequently used as an integrator of the differential system. We have shown the extreme robustness of our difference scheme, by performing a simulation at the limit where the time step can be taken arbitrarily large. 

The model focused mainly on the impact of various lockdown-exit strategies on the evolution of the epidemic. However, before addressing this question, we studied the epidemic onset and its evolution in the case where a substantial nucleus of infected persons does exists compared to the case where one starts with a single infected parson. It turned out that the observed evolution was the same in the two cases. The practical implication of this is that one cannot tell when did the first infections made their appearance. This observation resonates with the speculation [\refdef\reuters] that there may have been COVID-19 cases in Italy as early as November 2019. 

In order to apply our model in an as realistic as possible way we started by calibrating it and choosing properly initial conditions.
We studied different scenarios of lifting confinement: an abrupt one, a progressive one (where the infection rate of confined people goes back to the value of the unconfined linearly), and a stepwise one (where the infection rate of confined people takes an intermediate value, between the confined and the unconfined values, before going back to the unconfined value). 
First, we show that an abrupt end of confinement can lead to a second peak of epidemic higher than the first one, for a confinement that lasts 10 units of time in the model (50 days). 
However, even with an abrupt end, the ratio between the two peaks of epidemic can be smaller than 1, but this requires a long confinement, longer than 12.6 units of time (which represents 63 days), see Figure 11. 
In practice, in order to be sure that the second peak is much lower than the first one, the strict lockdown should be imposed for at least 2.5-3 months. Practically, however, it would be difficult to impose to the population a strict lockdown during such a long time interval.

A more acceptable strategy would be to alleviate the constrains progressively or stepwise. We studied both scenarios. The progressive scenario can be described by two variables: the time interval corresponding to a strict lockdown ($T_1$) and the time of progressive exit of the lockdown ($T_2$).  It is not surprising to find that the ratio of the two peaks is smaller when the lifting of constrains starts later (large $T_1$). For a given $T_1$, the ratio is smaller when the time of progressive exit of confinement ($T_2$) increases (going from curve (a) to (b), (c) and (d) on a vertical line, on Figure 11). 

What is less intuitive is that the total period of confinement (including the strict and the lighter confinement) has to be longer in the case of the linear exit than for the abrupt exit, in order to get a second peak lower than the first one. On figure 11, for curve (d) for example (that corresponds to $T_2$=7), the time interval for strict confinement $T_1$ has to be equal to 8 units of time if we want a ratio smaller than 1. The total time of confinement (including the period of strict and of lighter confinement) would thus be 15 units of time. For the abrupt exit, a duration of confinement of 12.5 units of time would suffice in order to have a ratio smaller than 1. 

Figure 11 also tells us that first, the ratios for an abrupt exit and for a progressive one tend to the same (lower than 1) value, as the time of confinement $T_1$ gets larger. With $T_1=30$ (meaning that the confinement would last 150 days, or 5 months), the number of infected people gets really small, but since it is not zero, the epidemics can start again when the lockdown is finally lifted, giving rise to a second, low peak, which has the same height whether the exit is progressive or not. We should point out at this point that applying a linear exit strategy in practice presents considerable difficulties. Changing rules every week or even everyday to ensure the linear lifting of the lockdown is far from ideal.

The stepwise strategy also allows to get ratios smaller than 1, if the duration of the confinement is long enough and if the intermediate infection rate $a_1$ for the confined people is not too large. However, what is interesting is that this intermediate value can be larger than 1: this means that even if the epidemic threshold is exceeded for the confined people, the peak of the epidemics can still be controlled, if the strict confinement period is long enough. For example, suppose that the confinement lasts 10 units of time in the model, or 50 days, (this situation corresponds to curve (c) in Figure 12), then any value of $a_1$ (the intermediate value of the infection rate of confined people) smaller than 1.8 would lead to a second peak lower than the first one. Moreover, this strategy allows to get less infected people at the end (see Figure 10).

This strategy seems to be easier to put into practice, as it does not necessitate a continuous change the rules. The difficult part here is to translate  the parameters of the model into practical instructions. What can represent the value of 1.8 for $a_1$? A situation between the strict confinement and a completely normal one, where, for example, everybody wears a mask, where those who can go to work do so, but work-place cafeterias are closed, where schools are still closed and where meetings of more than 10 persons are forbidden could be a good intermediate state. A more detailed model, with  a structured population would be needed to answer that question. However, in all cases one should bear in mind the words of Ferguson [\refdef\nature]: ``Models are not crystal balls. What we are building are simplified representations of reality''. Thus is some cases it is preferable to work with a simpler model which allows control over the various quantities in play and makes it possible to develop an intuitive grasp of the situation.

\bigskip
 {\scap References}
\medskip

%\begin{description}
\item{[\danbern]} D. Bernoulli, {\sl Essai d'une nouvelle analyse de la mortalit\'e caus\'ee par la petite v\'erole, et des avantages de l'inoculation pour la pr\'evenir}, Hist. et M\'em. de l'Acad. Roy. Sci. de Paris (1766) 1.
\item{[\kuchar]} A. Kucharski, T. Russell, C. Diamond, Y. Liu, J. Edmunds, S. Funk and R. Eggo, {\sl Early dynamics of transmission and control of COVID-19: a mathematical modelling study}, to appear in Lancet Infect Dis 2020 (https://doi.org/10.1016/ S1473-3099(20)30144-4).
\item{[\hellew]} J. Hellewell, S. Abbott, A. Gimma, N. Bosse, C. Jarvis, T. Russell, J. Munday, A. Kucharski, J. Edmunds, S. Funk and
R. Eggo, {\sl Feasibility of controlling COVID-19 outbreaks by isolation of cases and contacts}, Lancet Glob Health  8 (2020) e488. 
\item{[\klepac]} P. Klepac, A. Kucharski, A. Conlan, S. Kissler, M. Tang, H. Fry, and J. Gog, {\sl Contacts in context: large-scale setting-specific social mixing matrices from the BBC Pandemic project}, 
\hfill\break medRxiv preprint doi: https://doi.org/10.1101/2020.02.16.20023754.
\item{[\prem]} K. Prem, Y. Liu, T. Russell, A. Kucharski, R. Eggo, N. Davies, M. Jit and P. Klepac, {\sl The effect of control strategies to reduce social mixing on outcomes of the COVID-19 epidemic in Wuhan, China: a modelling study}, to appear in Lancet Public Health (2020) (https://doi.org/10.1016/ S2468-2667(20)30073-6).
\item{[\imper]} N.M. Ferguson and the Imperial College COVID-19 Response Team, {\sl Impact of non-pharmaceutical interventions (NPIs) to reduce COVID- 19 mortality and healthcare demand},  preprint March 2020.
\item{[\mordec]} E. Mordecai, M. Childs, M. Kain, N. Nova, J. Ritchie and M. Harris, {\sl Potential Long-Term Intervention Strategies for COVID-19}, online at https://covid-measures.github.io.
\item{[\gupta]} J. Louren\c{c}o, R. Paton, M. Ghafari, M. Kraemer, C. Thompson, P. Simmonds, P. Klenerman and S. Gupta, {\sl Fundamental principles of epidemic spread highlight the immediate need for large-scale serological surveys to assess the stage of the SARS-CoV-2 epidemic}, medRxiv preprint doi: https://doi.org/10.1101/2020.03.24.20042291.
\item{[\kermac]} W.O. Kermack, A.G. McKendrick, {\sl Contributions to the mathematical theory of epidemics},  Proc. Roy. Soc. Edinburgh A 115 (1927) 700.
\item{[\ross]} R. Ross, {\sl An application of the theory of probabilities to the study of a priori pathometry}, Proc. R. Soc. Lond. A  92 (1916) 204.
\item{[\extend]} J. Satsuma, R. Willox, A. Ramani, A.S. Carstea and B. Grammaticos, {\sl Extending the SIR epidemic model}, Physica A 336 (2004) 369.
\item{[\hethcot]} H. Hethcote, {\sl The Mathematics of Infectious Diseases}, SIAM Review 42 (2000) 599.
\item{[\bauch]} C. Bauch, D. Earn and M. Hillerman, {\sl Vaccination and the theory of games}, PNAS 101 (2004) 13391.
\item{[\mikens]} R.E. Mickens, {\sl Exact solutions to a finite-difference model of a nonlinear reaction-advection equation: Implications for numerical analysis},  Numer. Methods Partial Diff. Eq. 5 (1989) 313.
\item{[\handy]}  B. Grammaticos, R. Willox and J. Satsuma, {\sl Revisiting the Human and Nature Dynamics model}, Reg. Chao. Dyn. 25 (2020) 178.
\item{[\li]}  Q. Li, X. Guan, P. Wu, X. Wang, L. Zou, Y. Tong et al, {\sl Early Transmission Dynamics in Wuhan, China, of Novel Coronavirus–Infected Pneumonia}, New England J. of Med. 382 (2020) 1199.
\item{[\zhao]}  S. Zhao, Q. Lin, J. Ran, S. S. Musa, G. Yang, W. Wang et al, {\sl Preliminary estimation of the basic reproduction number of novel coronavirus (2019-nCoV) in China, from 2019 to 2020: A data-driven analysis in the early phase of the outbreak}, Int. J. of Inf. Diseases 92 (2020) 214.
\item{[\italy]} Data from https://www.worldometers.info/coronavirus/country/italy/, April 16th, 2020.
\item{[\lili]}  R. Li, S. Pei, B. Chen, Y. Song, T. Zhang, W. Yang and J. Shaman, {\sl Substantial undocumented infection facilitates the rapid dissemination of novel coronavirus}, Science, 10.1126/science.abb3221 (2020).
\item{[\reuters]} E. Parodi and S. Aloisi, {\sl Italian scientists investigate possible earlier emergence of coronavirus} 
\hfill\break 
from:  www.reuters.com, article-idUSKBN21D2IG (2020).
\item{[\nature]}  D. Adam, {\sl Special report: The simulations driving the world's response to COVID-19}, 
\hfill\break in https://www.nature.com/articles/d41586-020-01003-6 (2020).
%\end{description}

\end